\documentclass[%
 reprint,
superscriptaddress,
 amsmath,amssymb,
 aps,
prl,
]{revtex4-2}

\usepackage[
  colorlinks=true,
  linkcolor=blue,
  citecolor=blue,
  urlcolor=blue
]{hyperref}

\usepackage{graphicx}
\usepackage{dcolumn}
\usepackage{bm}

\begin{document}

\preprint{APS/123-QED}

\title{Symmetry-Empowered Through-Barrier Sensing in Complex Media}

\author{Shuai S. A. Yuan}
 \affiliation{Department of Electronics and Nanoengineering, Aalto University, 00076 Espoo, Finland}
\author{Zhazira Zhumabay}
\affiliation{Univ Rennes, CNRS, IETR - UMR 6164, F-35000, Rennes, France}%
 \author{Viktar Asadchy}%
 \affiliation{Department of Electronics and Nanoengineering, Aalto University, 00076 Espoo, Finland}
\author{Philipp del Hougne}
\email{philipp.del-hougne@univ-rennes.fr}
\affiliation{Department of Electronics and Nanoengineering, Aalto University, 00076 Espoo, Finland}%
\affiliation{Univ Rennes, CNRS, IETR - UMR 6164, F-35000, Rennes, France}%

\begin{abstract}
Symmetry strongly impacts wave transport in complex media. In this Letter, we demonstrate that the phenomenon of symmetry-induced through-barrier transmission enhancement enables quantitative sensing across barriers in complex media. We consider two mirror-symmetric chaotic cavities coupled through a narrow slit and containing point scatterers at mirror-symmetric positions. The characteristics of the scatterers in one cavity are unknown, whereas those of the scatterers in the other cavity are programmable. By tuning the programmable scatterers to maximize broadband total transmission, we recover the unknown scatterers' characteristics across the barrier. We show that reliable sensing requires a sufficiently large bandwidth, because otherwise a narrowband asymmetric resonant enhancement can dominate over the desired symmetry-induced enhancement. We further examine how absorption and barrier opacity influence the minimum required bandwidth. Our results establish a symmetry-empowered principle for through-barrier sensing in complex media, suggesting a route toward through-wall imaging in complex environments.
\end{abstract}

\maketitle

Symmetry principles play a fundamental role in our understanding and control of many wave phenomena. In particular, the presence or absence of symmetries can enable extreme wave control in tailored artificial metamaterials~\cite{yves2025symmetry}. For wave transport in complex and disordered systems~\cite{rotter2017light,cao2022shaping}, symmetry-induced coherent interference phenomena can result in remarkable deviations from diffusion theory. Of long-standing interest are complex wave systems with spatial symmetries~\cite{baranger1996reflection,whitney2009semiclassical_I,whitney2009semiclassical_II,gopar2006transport,kopp2008staggered,birchall2012random,gorbatsevich2016pt,gorbatsevich2017coalescence,gorbatsevich2017unified,dhia2018trapped,sweeney2020theory,stone2020reflectionless,ferise2022exceptional,saini2024mirror,jiang2024coherent,zhumabay2026inverse}. In this Letter, our focus is on complex wave systems with spatial left-right symmetry, where a barrier is placed at the interface between the two symmetric system halves. In the context of electron transport through quantum dots, a significant conductance enhancement through a tunnel barrier was demonstrated in symmetric double quantum dots of chaotic shape~\cite{macucci2007tunneling,whitney2009huge,totaro2010effect,macucci2020optimization}. Similar observations were made for graphene-based double dots~\cite{marconcini2013symmetry}. An analogous transmission enhancement was furthermore observed through opaque barriers sandwiched between symmetric diffusive disordered slabs~\cite{cheron2019broadband,cheron2020broadband,cheron2020sensitivity,davy2021experimental}. These symmetry-induced transmission enhancements through barriers are non-resonant, broadband effects because they originate from the constructive interference between pairs of multiple-scattering paths that acquire matching phases due to the left-right mirror symmetry~\cite{whitney2009huge,cheron2019broadband}. This physical picture is further supported by recent theoretical studies~\cite{borcea2024enhanced,flegontov2025symmetry}.

Naturally, the symmetry-induced transmission enhancement is quite sensitive to defects in the symmetry~\cite{whitney2009huge,totaro2010effect,macucci2020optimization,cheron2020broadband,cheron2020sensitivity}. Clearly, this sensitivity could be leveraged for defect detection~\cite{whitney2009huge,cheron2019broadband}. Besides such qualitative approaches, initial explorations on leveraging this sensitivity to quantitatively characterize a single-parameter perturbation (specifically, a magnetic field strength) were reported~\cite{macucci2007tunneling,macucci2020optimization}. However, it remains elusive to accurately reconstruct a multi-parameter perturbation in one half of a barrier-separated complex wave system based on the phenomenon of symmetry-induced transmission enhancement. Such a possibility would point toward novel symmetry-empowered principles for through-barrier sensing and, potentially, through-wall imaging in complex environments.

In this Letter, we study this possibility in the context of deterministically programmable complex wave systems. Specifically, we consider two mirror-symmetric chaotic cavities coupled via a thin slit and populated by point-like scatterers at mirror-symmetric locations. The scatterers in one cavity have static but unknown scattering characteristics, which we seek to infer by tuning the programmable scatterers in the other cavity so as to maximize the broadband total transmission. We show that successful sensing requires a sufficiently large bandwidth, because otherwise narrowband asymmetric resonant enhancements can exceed the desired broadband nonresonant enhancement induced by mirror symmetry. We further examine how the absorption strength and the barrier opacity (controlled by the slit width) affect the minimum bandwidth required for reliable sensing. Our work can be understood as realizing a discrete-basis form of symmetry-empowered through-barrier imaging.

\begin{figure}[t]
	\begin{center}
    \includegraphics [width=\columnwidth] {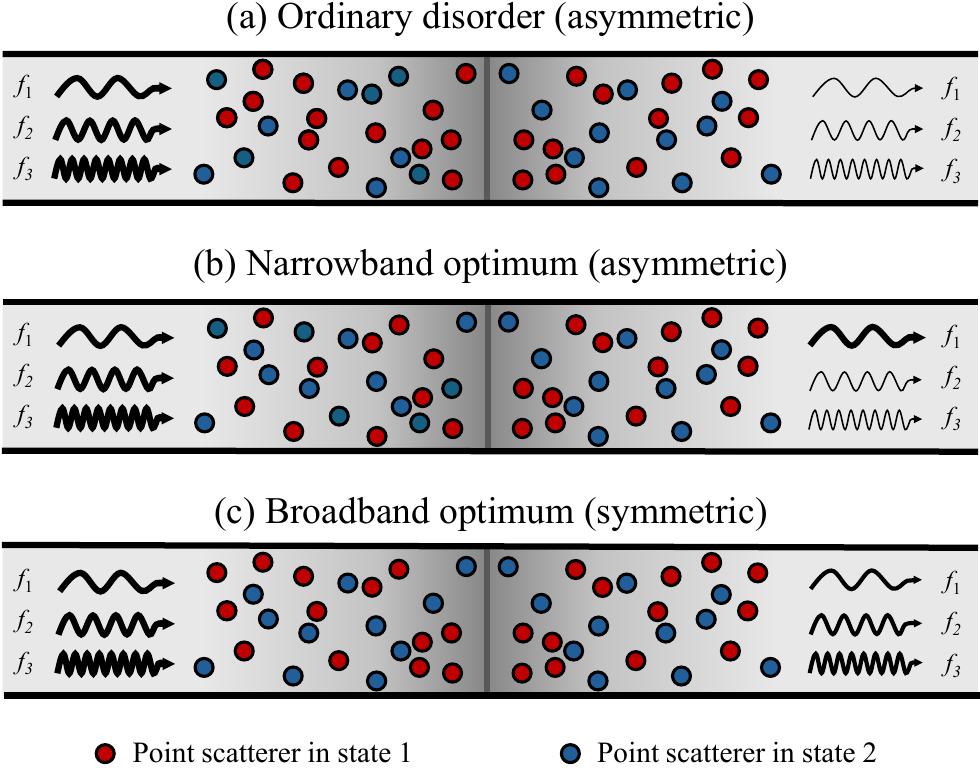}
    \caption{Transmission of different uncorrelated frequencies ($f_1$, $f_2$, $f_3$) through a barrier sandwiched between two slabs with mirror-symmetric point scatterers whose scattering characteristics can take two possible values (blue or red). Period and line thickness of the wavy arrows indicate the corresponding frequency and total transmitted power, respectively.  
    (a) For a randomly chosen configuration, the total transmitted power is low at all frequencies. 
    (b) When the left configuration is chosen to maximize the transmitted power over a narrow bandwidth, the optimum configuration is typically asymmetric and produces a strong resonant enhancement only over a limited spectral interval. 
    (c) When the left configuration is chosen to maximize the transmitted power over a broad bandwidth, the optimum configuration is mirror-symmetric and yields a robust broadband transmission enhancement. Thus, by optimizing the left configuration to maximize transmission over a sufficiently large bandwidth, one can identify the unknown right configuration through mirror symmetry.}
    \label{Fig1}
    	\end{center}
\end{figure}

To introduce the physical ingredients underpinning our approach, Fig.~\ref{Fig1} shows a schematic generic multimode waveguide with point scatterers on both sides of a barrier. The scatterer positions are random but mirror-symmetric with respect to the barrier, while asymmetry is introduced through their scattering characteristics. This differs from Ref.~\cite{cheron2019broadband}, where disorder was realized through the scatterers' positions rather than through their scattering characteristics. This deviation from Ref.~\cite{cheron2019broadband} is deliberate to match the specific setup that we study below, where it affords convenient modeling, but the underlying symmetry-based principle is independent of how the asymmetry is physically realized. 

We denote by $\mathbf{T}(f)$ the $M \times M$ transmission matrix from the left side to the right side at frequency $f$, where $M$ is the number of channels coupled to the system on either side. In the waveguide setup sketched in Fig.~\ref{Fig1}, these channels correspond to waveguide modes; in the specific setup we study below, they will correspond to discrete point sources and point detectors. Next, we define the total transmitted power metric $\tau(f) = \mathrm{Tr}\!\left[\mathbf{T}(f)\mathbf{T}^\dagger(f)\right]$; Ref.~\cite{cheron2019broadband} refers to $\tau(f)$ as the conductance. 
We denote the choice of the states of the scatterers (on the left and right sides) as the system's (left and right) ``configuration'' throughout this Letter. We assume the right configuration is fixed and unknown, while the left configuration is tunable and known. Ultimately, our goal is to identify the unknown right configuration based on how $\tau(f)$ depends on the left configuration.

With ordinary (asymmetric) disorder in Fig.~\ref{Fig1}a, $\tau(f)$ is generally low at all frequencies due to the barrier. With symmetric disorder in Fig.~\ref{Fig1}c, one can observe a broadband non-resonant enhancement of $\tau(f)$~\cite{whitney2009huge,cheron2019broadband}. In this Letter, we ask whether one can thus identify the right configuration by optimizing the left configuration to maximize the total transmitted power. A prerequisite for such a symmetry-based sensing protocol is that there is no asymmetric configuration that yields a larger enhancement of $\tau(f)$ than the symmetric configuration. 

To satisfy this requirement, we must reap the broadband property of the symmetry-enhanced through-barrier transmission. Indeed, if we maximize the bandwidth-averaged total transmitted power $\mathcal{T}(f_0,\Delta f)=\frac{1}{\Delta f}\int_{f_0 - \Delta f/2}^{f_0 + \Delta f/2} \tau(f) \,\mathrm{d}f$ only for a narrow frequency interval $\Delta f$ around a central frequency $f_0$, then it is very likely that the optimal configuration is asymmetric with its $\mathcal{T}_\mathrm{opt}(f_0,\Delta f)$ exceeding $\mathcal{T}_\mathrm{sym}(f_0,\Delta f)$ achieved by the symmetric configuration. Such a resonant enhancement with an optimal asymmetric configuration, illustrated in Fig.~\ref{Fig1}b, is spectrally confined to frequencies within $f_0 - (\Delta f + \delta f_\mathrm{corr})/2<f<f_0 + (\Delta f + \delta f_\mathrm{corr})/2$, where the correlation frequency interval $\delta f_\mathrm{corr}$ is the minimal spectral length guaranteeing uncorrelated frequencies~\cite{derode2001random_I,derode2001random_II}. 
Frequencies further away from $f_0$ are generally not expected to experience any enhancement. Selecting a large $\Delta f$ is thus the key to avoiding that such a resonant phenomenon can outperform the non-resonant symmetry-based enhancement. 

\begin{figure}[ht]
	\begin{center}
\includegraphics [width=\columnwidth] {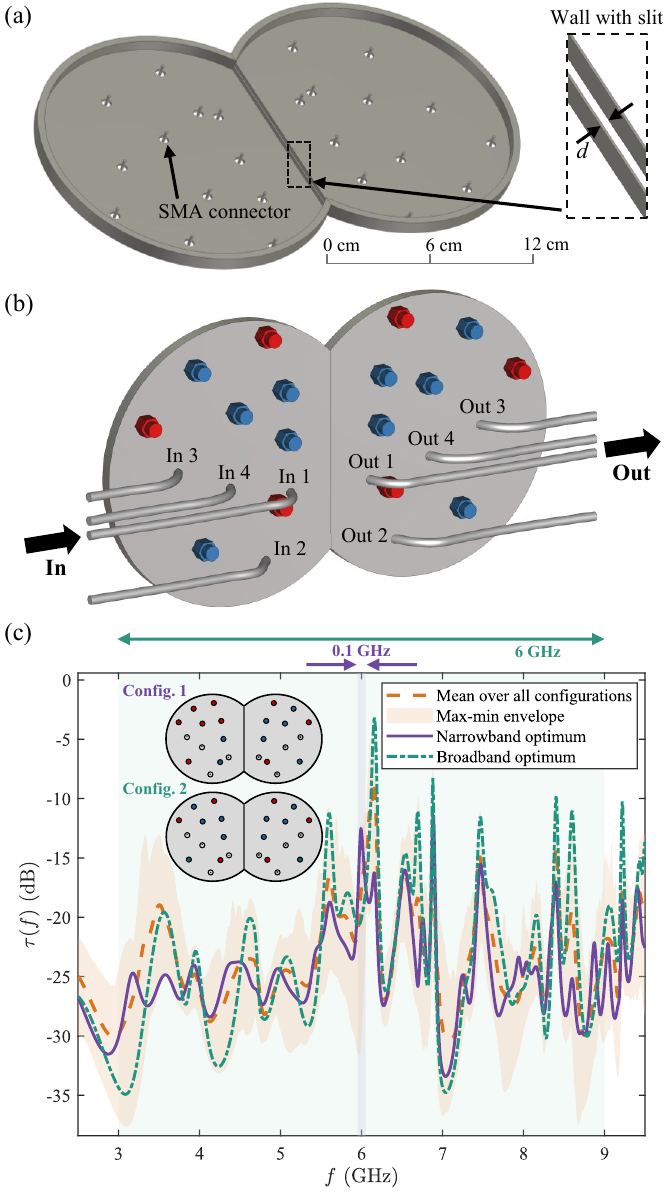}
	\caption{(a,b) Considered setup viewed from above with the cover removed to show the interior in (a), and from below in (b), where the tunable loads (blue and red) and coaxial cables are visible. (c) Spectra of $\tau(f)$ for a left configuration optimized for a narrow 0.1~GHz interval (purple) or for a broad 6~GHz interval (green). The corresponding configurations are shown as inset; the broadband optimal configuration is symmetric. In addition, we plot the mean of $\tau(f)$ across all left configurations and the corresponding max-min envelope. }
	\label{Fig2}
	\end{center}
\end{figure}

To study our proposed symmetry-empowered sensing concept in detail, we consider the setting displayed in Fig.~\ref{Fig2}a. We make two important choices. First, as mentioned, unlike Ref.~\cite{cheron2019broadband} we fix the point scatterers' positions and instead tune their scattering characteristics. Thereby, a single full-wave simulation is sufficient, and subsequently we can leverage a computationally efficient closed-form model from multiport network theory (MNT) to predict $\mathbf{T}(f)$ for any desired configuration~\cite{tapie2024systematic}. We summarize details of this MNT technique in the End Matter. Second, we limit the number of scatterers on either side to $P=8$ and the number of possible states per scatterer to two. Thereby, we can evaluate all $2^8=256$ possible left configurations (for a fixed right configuration) and identify the globally optimal one for maximizing $\mathcal{T}(f_0,\Delta f)$. We can thus focus on the physics without worrying about whether an optimization algorithm identifies the global optimum. As seen in Fig.~\ref{Fig2}a, the scatterers on the left side and right side of the barrier are embedded in mirror-symmetric, reciprocal, quasi-2D D-shaped copper cavities of height 4~mm. Such D-shaped cavities are known to give rise to wave chaos and thus a widely used example of complex media~\cite{draeger1997one,ree1999classical,doya2001optimized,redding2015low,faul2025agile}. The barrier is realized by the common separating wall between the two cavities, except for a narrow slit of width $d$ opened at mid-height. There are $M=4$ coherent point sources in the left cavity and $M=4$ coherent point detectors in mirror-symmetric positions in the right cavity. We note that Ref.~\cite{whitney2009huge} also considered a mirror-symmetric cavity with a barrier along the symmetry axis and symmetrically connected to multiple channels on either side. Our setup is mirror-symmetric when the left and right configurations of the point scatterers are identical.

As seen in Fig.~\ref{Fig2}a, the point sources and point detectors are realized as deeply subwavelength SubMiniature version A (SMA) connectors whose center pins protrude into the cavity and are connected to coaxial cables. Each coaxial cable is a single-mode waveguide constituting one scattering channel. The tunable point scatterers are also realized as SMA connectors inserted into the cavity at prescribed positions, but they are terminated externally either by open-circuit (OC) or short-circuit (SC) loads. This implementation is fully standard in microwave engineering and closely follows common experimental realizations of programmable wave-chaotic cavities and programmable metasurfaces~\cite{del2025experimental}.  

In Fig.~\ref{Fig2}c, we pick an arbitrary right configuration and display the spectra of $\tau(f)$ associated with two distinct left configurations. Configuration 1 is globally optimal to maximize $\mathcal{T}(6\ \mathrm{GHz},0.1\ \mathrm{GHz})$ while Configuration 2 is globally optimal to maximize $\mathcal{T}(6\ \mathrm{GHz},6\ \mathrm{GHz})$. Upon inspection of the two configurations, it is apparent that the latter is perfectly symmetric while the former is clearly not. The associated spectra are plotted in purple ($\Delta f = 0.1$~GHz) and green ($\Delta f = 6$~GHz) in Fig.~\ref{Fig2}c. In the vicinity of $f_0$, the purple spectrum displays a clear resonance that strongly enhances $\tau(f)$ beyond its mean value (orange dashed line), reaching $\mathcal{T}_\mathrm{1}(6\ \mathrm{GHz},0.1\ \mathrm{GHz})=3.39\times10^{-2}$. Meanwhile, the green spectrum is even below the mean because it seemingly has no resonance close to $f_0$, only reaching $\mathcal{T}_\mathrm{2}(6\ \mathrm{GHz},0.1\ \mathrm{GHz})=1.09\times10^{-2}$. This observation corroborates that for small $\Delta f$ the symmetric configuration is by no means guaranteed to be optimal in terms of maximizing $\mathcal{T}(f_0,\Delta f)$ because the asymmetric Configuration 1 achieves a strong resonant enhancement. That resonant enhancement is, however, by its very nature spectrally confined. Within the broad 6~GHz interval for which the Configuration 2 is optimized, Configuration 1 only achieves $\mathcal{T}_\mathrm{1}(6\ \mathrm{GHz},6\ \mathrm{GHz})=5.39\times10^{-3}$ while Configuration 2 achieves $\mathcal{T}_\mathrm{2}(6\ \mathrm{GHz},6\ \mathrm{GHz})=1.36\times10^{-2}$. For the specific considered right configuration, our results in Fig.~\ref{Fig2}c thus confirm the feasibility of symmetry-empowered sensing and emphasize the importance of a sufficiently large $\Delta f$.

\begin{figure}[t]
	\begin{center}
\includegraphics [width=\columnwidth] {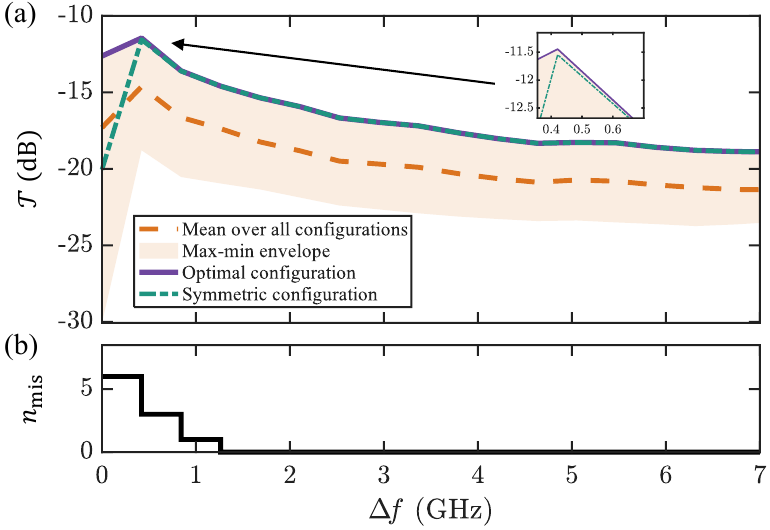}
	\caption{Dependence on $\Delta f$, for the right configuration seen in Fig.~\ref{Fig2}c and $f_0=6$~GHz. (a) Bandwidth-averaged total transmitted power $\mathcal{T}(f_0,\Delta f)$ for the optimal left configuration, the symmetric left configuration, and the mean over all left configurations. The optimal left configuration is determined separately for each value of $\Delta f$. (b) $n_\mathrm{mis}$ (number of incorrectly identified state variables).}
	\label{Fig3}
	\end{center}
\end{figure}

To systematically identify the minimum required value of $\Delta f$, we now examine how both $\mathcal{T}(f_0,\Delta f)$ and the mismatch count $n_\mathrm{mis}$ depend on $\Delta f$. We define $n_\mathrm{mis}$ as the number of state variables for which the optimal left configuration differs from the mirror image of the unknown right configuration, i.e., the number of incorrectly identified loads. As shown in Fig.~\ref{Fig3}a, in the narrowband limit the symmetric configuration is inferior to the globally optimal one, consistent with the resonant asymmetric enhancement already observed in Fig.~\ref{Fig2}c. However, $\mathcal{T}_\mathrm{sym}(f_0,\Delta f)$ of the symmetric configuration rapidly approaches $\mathcal{T}_\mathrm{opt}(f_0,\Delta f)$ of the global optimum (evaluated separately per value of $\Delta f$) as $\Delta f$ increases, indicating that the broadband nonresonant symmetry-induced enhancement rapidly dominates over narrowband resonant enhancements. Yet, sensing succeeds only once the symmetric configuration becomes the global optimum, because only then does the optimal left configuration coincide with the mirror image of the right one. For values of $\Delta f$ around 1~GHz, $\mathcal{T}_\mathrm{sym}$ is still slightly inferior to $\mathcal{T}_\mathrm{opt}$, meaning that there is a non-symmetric configuration that slightly outperforms the symmetric one and thereby prevents our sensing scheme from working. However, for values of $\Delta f$ above  roughly 1.3~GHz, the symmetric configuration is always the globally optimal one to maximize $\mathcal{T}$, as confirmed by $n_\mathrm{mis}=0$ in Fig.~\ref{Fig3}b.

\begin{figure}[ht]
	\begin{center}
    \includegraphics [width=\columnwidth] {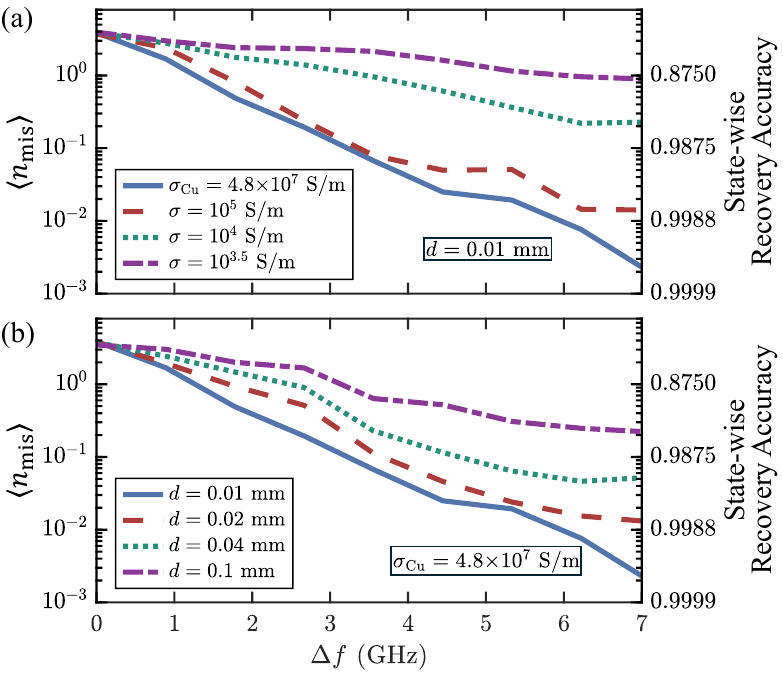}
    \caption{Bandwidth dependence of the average mismatch count $\langle n_\mathrm{mis}\rangle$, evaluated over all $2^8=256$ right configurations for 20 different choices of which four of the 12 mirror-symmetric SMA-connector pairs act as source/detector pairs. The right axes show the corresponding state-wise recovery accuracy, $1-\langle n_\mathrm{mis}\rangle/8$. (a) Effect of Ohmic attenuation, controlled by the cavity conductivity $\sigma$. (b) Effect of barrier opacity, controlled by the slit width $d$.}
    \label{Fig4}
    \end{center}
\end{figure}

So far, we only considered one arbitrarily chosen right configuration. To establish the generality of our result (i.e., to rule out that the right configuration considered in Fig.~\ref{Fig2}c and Fig.~\ref{Fig3} was a lucky coincidence), we now repeat the analysis underlying Fig.~\ref{Fig3} for all possible $2^8 = 256$ right configurations. 
In addition, we repeat the exhaustive analysis for 20 different choices of which four out of the 12 mirror-symmetric SMA-connector pairs act as point sources/detectors, i.e., are connected to scattering channels, while the remaining eight pairs act as tunable point scatterers terminated by tunable loads. 
Importantly, all these $20\times256=5120$ realizations are evaluated based on a single full-wave simulation, as detailed in the End Matter. We report the average of the mismatch count $\langle n_\mathrm{mis}\rangle$ across the $5120$ realizations as solid blue line in Fig.~\ref{Fig4}. We observe that increases of $\Delta f$ systematically lead to reductions of $\langle n_\mathrm{mis} \rangle$. For $\Delta f = 7\ \mathrm{GHz}$, $\langle n_\mathrm{mis}\rangle=2.7\times10^{-3}$, corresponding to only 14 incorrectly identified point scatterer states across the $20\times256=5120$ test cases, i.e., a state-wise recovery accuracy of $99.97\%$. Our sensing protocol thus becomes essentially deterministic and highly reliable once the bandwidth is large compared to the spectral scale over which accidental resonant enhancements persist.

Finally, we seek to understand the qualitative influence of attenuation and barrier opacity on the bandwidth required by our sensing protocol. In Fig.~\ref{Fig4}a, we systematically reduce the cavity conductivity $\sigma$, thereby increasing Ohmic attenuation. We observe that stronger attenuation leads to a slower decay of $\langle n_\mathrm{mis}\rangle$ with $\Delta f$, and therefore to a larger bandwidth required for reliable sensing. This trend is physically intuitive because the relevant quantity is not the absolute bandwidth $\Delta f$, but rather the number of independent spectral degrees of freedom contained in it, which scales approximately as $\Delta f/\delta f_\mathrm{corr}$. Stronger attenuation broadens the spectral correlation width $\delta f_\mathrm{corr}$, so that a given $\Delta f$ spans fewer independent spectral samples. Consequently, a larger bandwidth is needed before the broadband symmetry-induced enhancement can reliably dominate over narrowband resonant enhancements. A complementary interpretation builds on the well-documented weakening of the symmetry-induced enhancement by attenuation, which suppresses long multiple-scattering paths~\cite{cheron2020sensitivity,flegontov2025symmetry}. As a result, the symmetric configuration is less clearly separated from competing asymmetric configurations, and its unambiguous dominance requires a larger bandwidth.

In Fig.~\ref{Fig4}b, we systematically reduce the barrier opacity by increasing the slit width $d$. We observe that reducing the barrier opacity leads to a slower decay of $\langle n_\mathrm{mis}\rangle$ with $\Delta f$, and therefore also increases the bandwidth required for reliable sensing. This trend is consistent with previous studies showing that the symmetry-induced enhancement  weakens as the bare barrier opacity is reduced~\cite{cheron2019broadband,borcea2024enhanced,flegontov2025symmetry}. Thus, for a less opaque barrier, the symmetric configuration is less clearly separated from competing asymmetric configurations. As in the case of increased attenuation, its unambiguous dominance over narrowband resonant enhancements therefore requires a larger number of independent spectral samples, and hence a larger bandwidth.

To summarize, we have demonstrated a symmetry-empowered principle for quantitative through-barrier sensing in complex media.  In a mirror-symmetric chaotic-cavity system, we recovered an unknown binary configuration of point scatterers on one side of an opaque barrier by tuning programmable scatterers on the other side to maximize bandwidth-averaged total transmission. Reliable recovery requires a sufficiently large bandwidth because narrowband optimizations can prefer resonances arising in asymmetric configurations, whereas broadband optimization favors the nonresonant symmetry-induced transmission enhancement. We achieved an accuracy of $99.97\%$ in the default case and found that stronger attenuation and lower barrier opacity both increase the bandwidth required for reliable sensing.

Looking forward, the proposed protocol could be implemented either directly in situ, by optimizing programmable scatterers using measured transmission data, or indirectly in software based on an experimentally calibrated MNT system model. The former case could avoid the need for phase-resolved measurements by relying on short-pulse excitation and maximizing the time- and space-integrated transmitted energy as a proxy for $\mathcal{T}$. The latter case could capitalize on recent progress in experimental MNT parameter estimation for programmable metasurfaces~\cite{sol2024experimentally,del2025experimental,del2026ambiguity,del2026reduced,tapie2025experimental,tapie2026channel}. For larger systems, where exhaustive search over all configurations is infeasible, future work can combine scalable optimization strategies with recently developed electromagnetic bounds on the achievable total transmission in programmable MIMO systems~\cite{del2026electromagnetic} to probe how close an optimized configuration is to the global optimum.

\begin{acknowledgments}
The authors acknowledge fruitful discussions with Constantin Simovski and Sergei Tretyakov. 
This work was supported in part by the Nokia Foundation (project 20260028), the ANR France 2030 program (project ANR-22-PEFT-0005), the ANR PRCI program (project ANR-22-CE93-0010), and the Research Council of Finland (projects 371367 and 365679).
\end{acknowledgments}

\providecommand{\noopsort}[1]{}\providecommand{\singleletter}[1]{#1}%

\section*{End Matter}

We summarize here how we predict $\mathbf{T}(f)$ for any desired configuration based on a single full-wave simulation. As mentioned in the main text, our point sources, point detectors, and tunable point scatterers are all realized based on deeply subwavelength SMA connectors. In microwave engineering, the deeply subwavelength SMA connectors are referred to as electrically small lumped ports, implying that the field variation over the port cross-section is negligible. 
Without the loads and the coaxial cables, our system is a perfectly static system with $N=2(M+P)=24$ ports whose scattering properties are fully characterized by the $N\times N$ scattering matrix $\mathbf{S}(f)$. We extract $\mathbf{S}(f)$ with a single full-wave simulation in the 3D electromagnetic simulation software CST Studio Suite~\cite{tapie2024systematic}. If an $N$-element coherent wavefront $\mathbf{a}(f)$ was injected via all $N$ ports, then the outgoing wavefront would be $\mathbf{b}(f) = \mathbf{S}(f)\,\mathbf{a}(f)$.

However, in our setup $2P$ ports are terminated.
We partition the $N$ ports into two disjoint sets: $\mathcal{A}$ contains the $2M$ ports connected to scattering channels, and $\mathcal{B}$ contains the $2P$ ports connected to tunable loads. This leads to the following partition of $\mathbf{S}$:
\[
\begin{pmatrix}
\mathbf{b}_\mathcal{A}\\
\mathbf{b}_\mathcal{B}
\end{pmatrix}
=
\begin{pmatrix}
\mathbf{S}_\mathcal{AA} & \mathbf{S}_\mathcal{AB}\\
\mathbf{S}_\mathcal{BA} & \mathbf{S}_\mathcal{BB}
\end{pmatrix}
\begin{pmatrix}
\mathbf{a}_\mathcal{A}\\
\mathbf{a}_\mathcal{B}
\end{pmatrix}.
\]
Here, $\mathbf{S}_{\mathcal{XY}}$ denotes the submatrix of $\mathbf{S}$ mapping incoming waves at the ports in the index set $\mathcal{Y}$ to outgoing waves at the ports in the index set $\mathcal{X}$, with $\mathcal{X},\mathcal{Y}\in\{\mathcal{A},\mathcal{B}\}$.

The tunable loads impose linear relations between the incoming and outgoing waves at the terminated ports,
\[
\mathbf{a}_\mathcal{B}(f)=\mathbf{\Gamma}(f)\,\mathbf{b}_\mathcal{B}(f),
\]
where $\mathbf{\Gamma}(f)$ is a diagonal $2P\times 2P$ matrix whose diagonal entries are the reflection coefficients of the chosen loads. For ideal OC and SC terminations,
\[
\mathbf{\Gamma}(f)=\mathrm{diag}(r_1,\dots,r_{2P}), \qquad r_i\in\{+1,-1\},
\]
where $r_i=+1$ and $r_i=-1$ correspond to OC and SC, respectively. Eliminating $\mathbf{a}_\mathcal{B}$ and $\mathbf{b}_\mathcal{B}$ yields an effective reduced scattering matrix for the accessible ports,
\[
\mathbf{b}_\mathcal{A}(f)=\tilde{\mathbf{S}}(f)\,\mathbf{a}_\mathcal{A}(f),
\]
with
\[
\tilde{\mathbf{S}}(f) 
= 
\mathbf{S}_\mathcal{AA}(f)
+
\mathbf{S}_\mathcal{AB}(f)\, \mathbf{\Phi}(f) \,
\mathbf{S}_\mathcal{BA}(f),
\]
where $\mathbf{\Phi}(f)=\mathbf{\Gamma}(f)
\left[\mathbf{I}-\mathbf{S}_\mathcal{BB}(f)\mathbf{\Gamma}(f)\right]^{-1}$.
This standard terminated-port reduction formula from MNT self-consistently accounts for all multiple-scattering events involving the loaded ports in closed form. This can be seen more explicitly by rewriting the matrix inversion as Neumann series (see, for instance, Eq.~(3) in Ref.~\cite{del2025physics}).

In our setup, we further partition the $2M$ ports connected to scattering channels into the $M$ transmitting left ports and the $M$ receiving right ports. We refer the corresponding disjoint sets as $\mathcal{A}_\mathrm{L}$ and $\mathcal{A}_\mathrm{R}$, respectively, where $\mathcal{A}=\mathcal{A}_\mathrm{L}\cup\mathcal{A}_\mathrm{R}$. This partition allows us to identify
\[
\mathbf{b}_{\mathcal{A}_\mathrm{R}}(f)={\mathbf{T}}(f)\,\mathbf{a}_{\mathcal{A}_\mathrm{L}}(f),
\]
where $\mathbf{T}(f)$ is an off-diagonal block of $\tilde{\mathbf{S}}$ given by
\[
{\mathbf{T}}(f) 
= 
\mathbf{S}_{\mathcal{A}_\mathrm{R}\mathcal{A}_\mathrm{L}}(f)
+
\mathbf{S}_{\mathcal{A}_\mathrm{R}\mathcal{B}}(f)\,\mathbf{\Phi}(f) \,
\mathbf{S}_{\mathcal{B}\mathcal{A}_\mathrm{L}}(f).
\]

\end{document}